\documentclass[prl,twocolumn,letterpaper]{revtex4}

\usepackage{graphicx}
\usepackage{pslatex}

\begin{document}

\title{Ballooning Spiders: The Case for Electrostatic Flight}


\author{Peter W. Gorham}
\affiliation{ Dept. of Physics and Astronomy, Univ. of Hawaii, Manoa, HI 96822. }


\begin{abstract}
We consider general aspects of the physics underlying the flight of Gossamer spiders, also known as
ballooning spiders. We show that existing observations and the physics of spider silk in the
presence of the Earth's static atmospheric electric field indicate a potentially
important role for electrostatic forces in the flight of Gossamer spiders.
A compelling example is analyzed in detail, motivated by the observed 
``unaccountable rapidity'' in the launching
of such spiders from {\it H.M.S. Beagle}, recorded by Charles Darwin during his famous voyage.
\end{abstract}


\maketitle

Observations of the wide aerial dispersal of Gossamer spiders by kiting or ballooning
on silken threads have been described since
the mid-19th century~\cite{Blackwall1827,Murray1830,Darwin,McCook1877}.
Remarkably, there are still aspects of this behavior which remain in tension with
aerodynamic theories~\cite{Humphrey87,Suter99} in which 
the silk develops buoyancy through wind and convective
turbulence. Several observed aspects of
spider ballooning are difficult to explain in this manner: 
the fan shaped structures that multi-thread launches
employ~\cite{Darwin}, the capacity to
launch at surprisingly high initial (and even non-vertical) acceleration in conditions
where air movement is imperceptible~\cite{Darwin,Schneider01}; 
the ability of some relatively heavy adult spiders
to initiate ballooning~\cite{Schneider01}; and the surprising altitudes -- up to at least several km --
that are achieved by some ballooning groups~\cite{Coad31,Glick39}. 


It is curious that, given the large magnitude of the Earth's vertical
atmospheric electrostatic field,  there appears to be no
prior quantitative assessment of the possible action of
electrostatic buoyancy to provide a lift component separate from
purely aerodynamic effects. This field exists globally in the atmosphere 
with an average surface magnitude of
120 V m$^{-1}$ pointing downward~\cite{Bennett08}; the global charges required to sustain it
originate in charge separation in clouds and
electrically active storms~\cite{Fleagle80}.

The role of this electrostatic field in ballooning spider behavior remains unknown, despite several fascinating 
observations by no less a naturalist than
Charles Darwin in his voyage aboard {\it H.M.S. Beagle} from 1831-1836. 
Darwin writes in detail
of one particular period in the voyage, 60 miles off the coast of
Argentina, where the ship was inundated by ballooning spiders on a
relatively calm, clear day~\cite{Darwin}. 

Darwin observed distinct launching
behaviors in two separate species, including likely juveniles with
sizes in the range of 2-3 mm, and larger spiders of sizes around 7 mm. Of
the former, he reported that {\it ``I repeatedly observed the same kind of
small spider, either when placed or having crawled on some little eminence,
elevate its abdomen, send forth a thread, and then sail away horizontally,
but with a rapidity which was quite unaccountable.''} Of the latter spider,
Darwin writes that it, {\it  ``while standing on the summit 
of a post, darted forth four or five threads from its spinners. These,
glittering in the sunshine, might be compared to diverging rays of
light; they were not, however, straight, but in undulations like films
of silk blown by the wind. They were more than a yard in length,
and diverged in an ascending direction from the orifices. The spider
then suddenly let go of its hold on the post, and was quickly borne
out of sight. The day was hot and apparently quite calm...''} 

Darwin conjectured that imperceptible thermal convection of the
air might account for the rising of the web, but noted that the 
divergence of the threads in the latter case was likely to be
due to some electrostatic repulsion, a theory supported by
observations of Murray published in 1830~\cite{Murray1830}, but
earlier rejected by his contemporary Blackwall~\cite{Blackwall1827}.
The center of the controversy appears to be whether thermal convection
can account for the initial buoyancy of the threads as they are emitted;
recent detailed observations~\cite{Eberhard87} have not yet settled the
question of how the lines are spontaneously initiated under very calm conditions.

\begin{figure}[htb!]
\includegraphics[width=3.15in]{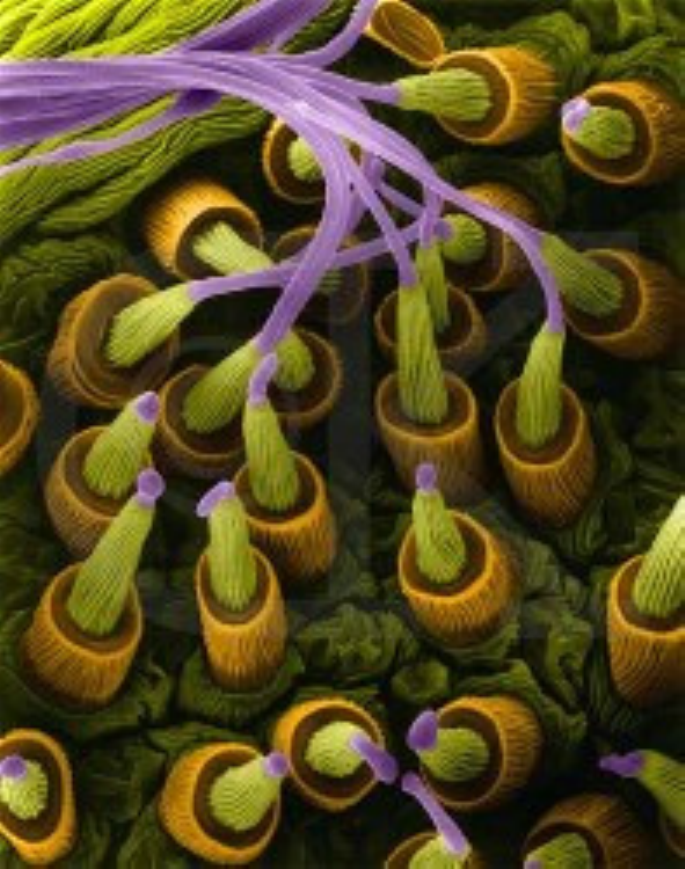}
\caption{Spiny-backed spider piriform gland spigots 
(from the anterior spinneret) through which silk is being secreted (Gasteracantha sp.).
Magnification: x490, Type:scanning electron micrograph. 
Image copyright Dennis Kunkel Microscopy, Inc., used with permission.\label{spinnerets}}
\end{figure}

Observations similar to those of Darwin and Murray may be found in a
surprising recent account by Schneider {\em et al.} of
the multi-thread launching of adult individuals of the relatively
large spider {\em Stegodyphus}~\cite{Schneider01}, with masses of
order 100~mg, well above the limit for 
such behavior based on parameters estimated from
aerodynamic theory~\cite{Henschel95,Suter91}. Schneider {\em et al.} note the
apparent electrostatic repulsion of the multiple threads, forming as
a result a triangular fan shape, with `` tens to hundred of threads.'' 
Scheider {\em et al.} also observed rapid vertical acceleration
of the spiders in calm conditions, up to altitudes of 30~m, where
they could no longer be visually tracked.

We study here the hypothesis that such spiders are able to emit threads that
are either pre-loaded with a static electric charge, or rapidly
gain it during or just after the strand-spinning and thread-weaving process. In either case, the presence 
of this charge will lead both to mutual repulsion among
the emitted threads, and an additional overall induced electrostatic force
on the spider, providing a component of lift that is independent
of convection or aerodynamic effects.


The complex protein structure of spider silk
includes  a high percentage, 7-9\% each of 
the charge-bearing amino acids glutamic acid 
and arginine~\cite{Saravanan06} which may be generated in
a charged state as part of the spinning process,
or may be utilized to facilitate attachment
of charge drawn up from the local launching surface
as the strands are spun from the sharp nozzles of the spinneret.
A generic example (from {\it Gasteracantha cancriformus}) 
showing typical nozzle geometry -- with tip diameters of of order 10 microns -- is
shown in Fig.~\ref{spinnerets}.

The average negative surface charge
density of the earth is about 6~nC~m$^{-2}$, and may be much higher
on the prominences which ballooning spiders choose to launch from.
Local surface charge may thus provide a source for the thread charge.
However, this may not be the only source;
a triboelectric mechanism that may be
important in charging of the strands during 
the initial strand emission and thread weaving at the spinneret is the
process known as flow electrification~\cite{FlowElect1,FlowElect2}, in which the
silk protein could undergo frictional charging as it 
passes through the spinneret nozzle.

A lower limit on the initial charge state of the
silk arises if we assume Murray was
correct in his hypothesis that the silk in multi-thread observations
carried enough charge to spread and buoy
the threads away from each other through electrostatic repulsion. 
We assume that the negative charge in each thread is concentrated
at the end of the silk, and that the silk mass is negligible,
so that simple electrostatic equilibrium in the atmospheric electric
field obtains. The upward-directed fan structure then forms as a result
of the force balance of the mutual thread repulsion opposing the
electrostatic buoyancy which tries to straighten the threads to
a common direction. 

Using this simple model, a 5-thread solution
for 1~m length threads fanning to about 1 radian 
gives $Q_{min} \simeq 1$~nC total, or 200 pC per thread~\cite{5strandModel}.
As spider silk is known to have high resistivity,
this estimate is conservative; a more complete model would use
a distributed charge along the threads.This quantity of charge appears
well within the range of triboelectric mechanisms; for example,
observations of triboelectrification of houseflies shows that they quickly charge up to levels of
order 50 pC within a short time by merely walking across a synthetic dielectric
surface~\cite{flycharge}, and honeybees are well known to achieve charges of 45 pC from
movement on beeswax~\cite{honeybees}.

\begin{figure}[htb!]
\includegraphics[width=3.7in]{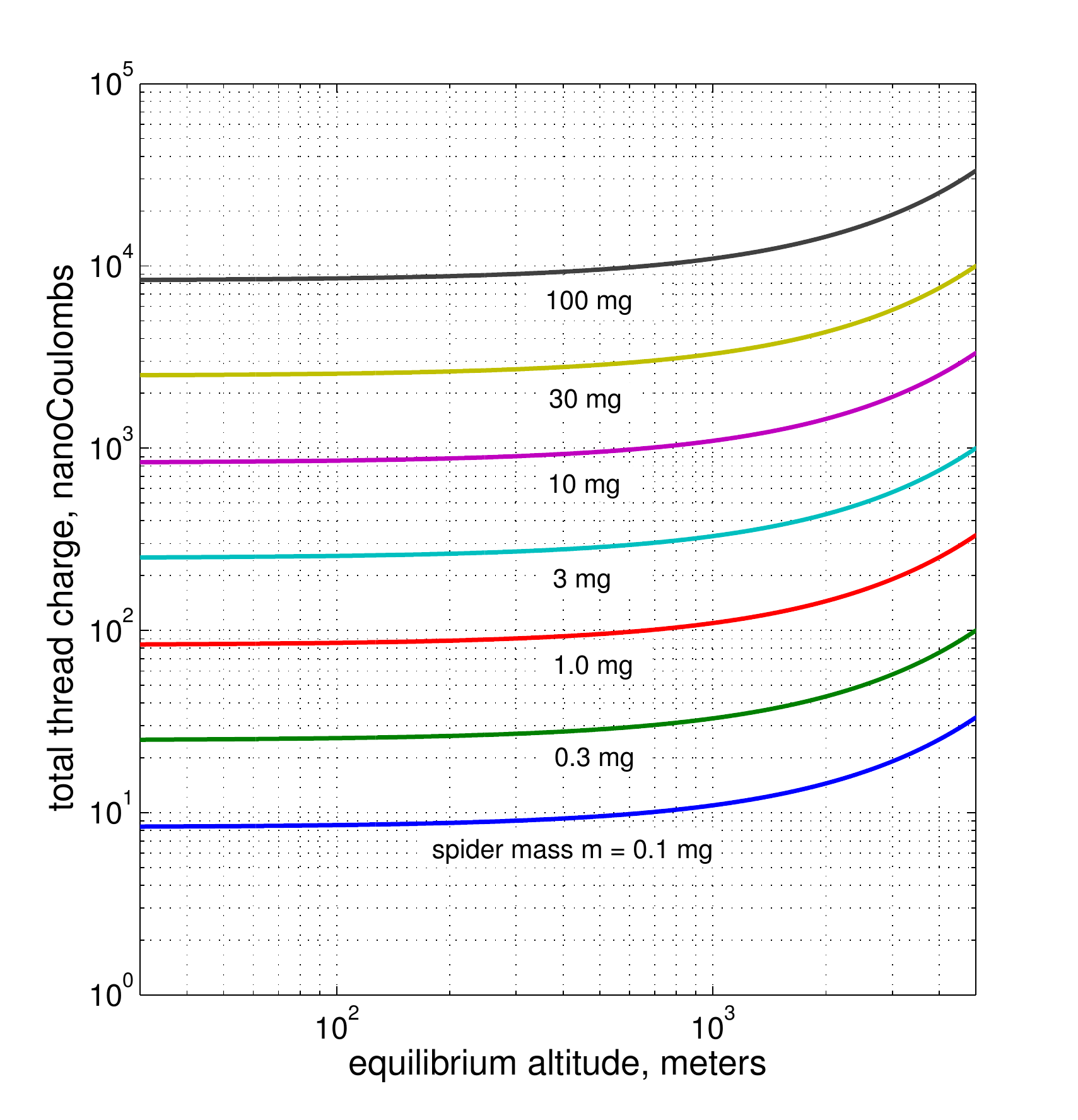}
\caption{Total charge (over all threads) required to reach the
equilibrium float altitudes indicated for spiders of various masses
shown in the labeled curves.\label{ChargevsH}}
\end{figure}

Unaided electrostatic flight will evidently require a much higher charge.
To assess the charge state required for this,
we use an approximate analytic model for the atmospheric electric field,
with an exponential fit to measured values at several altitudes~\cite{CRC};
a good approximation is given by
\begin{equation}
E(h) = E_0 e^{-\alpha h}~~{\rm V~m^{-1}}
\end{equation}
with $\alpha= 3.0 \times 10^{-4}$~m$^{-1}$ and $E_0 = -120$~V/m;
the integrated potential difference to an altitude of
50-60~km  is about 350KV. Ballooning spiders have been 
observed up to altitudes of order $4000$~m~\cite{Coad31};
an estimate of the total charge $Q$ of the thread can be obtained
by assuming that the float altitude represented the equilibrium
height $H_{eq}$ at which the weight of the spider was balanced by
the Coulomb lift force, that is
$ QE(H_{eq})= m g $
where $m$ is the spider mass, and $g=9.81$~m s$^{-2}$ is
the surface gravitational acceleration of the Earth.

We assume here that only a single thread is involved;
this appears to the the most common launch configuration.
Combining these equations,
\begin{equation}
Q = \frac{m g}{ E_0}~ e^{\alpha H_{eq}}.  
\end{equation}
For spiders of mass in the range of 0.1-0.3~mg, typical of 
ballooners~\cite{Coyle85}, the resulting $Q \simeq 10-30$nC (here we neglect
the mass of the silk lines, which amount to about 3~$\mu$g/m). 

Fig.~\ref{ChargevsH} shows a family of solutions for
various masses as a function of $H_{eq}$.
These values are a factor of 50 or more above the minimum
per thread required for developing the multithread fan.
They provide an 
upper limit for the charge state at launch; since it is likely that
both electrostatic and convective buoyancy will combine to create the lift needed
for flight, the initial charge state will be lower in proportion
to the aerodynamic contribution.

Given the much higher charge states required for electrostatic flight as compared
to buoyancy for a thread fan, the question of the source of the charge becomes more
acute. The charge may be intrinsic to the silk as it emerges from the spinneret,
through its complex protein chemistry and perhaps aided by the spinning process itself.

Electrification of various fluids and polymers through capillary flow 
has been extensively studied, but the details of the mechanism and its
application to a complex material such as spider silk, where internal
pressure and external tension both play a role in extruding the silk from
the spinneret, has not yet been quantified. Recent measurements of de-ionized water flow through capillaries
of several tens of microns diameter~\cite{CapFlowElect2012} give net streaming
currents of several femtoamperes per Pascal of nozzle pressure. 
Simple scaling of these data for somewhat smaller diameter of typical
spinneret nozzles, and an estimated extrusion pressure of order 10 kPa, 
the implied lower limit of the streaming current is of order
1 pA/strand for the same deonized water parameters. Since the charged boundary layer
in flow electrification depends on the Debye length in the extruding silk, which is
a function of the dielectric constant $\epsilon_r \simeq 10$ and conductivity 
$\sigma \simeq 4 \mu$S/m~\cite{ES_Thesis2012}, we can further scale these
results for the electrical properties of silk. Our scaling results indicate 
a possible streaming charge rate of order 3 nA per strand. Since a complete
silk thread may consist of 30 strands of more, and the thread may be spun out for
several seconds at least, a total charge per thread of tens of nC appears possible
by this process.

This is however not the only plausible charging mechanism;
charge may also be drawn up from the local surface charge in the vicinity of the
spider. When launching, spiders appear to strongly prefer prominences
above ground level; this behavior would be beneficial for both aerodynamic or
electrostatic accelerations. In the latter case, local charge distributions
become concentrated at convex prominences in the surface topography, 
and the electrostatic field will also become enhanced at such locations.

Assuming there was an adequate reservoir of surface charge available at a spider's 
launch site, we must determine
whether there is sufficient time for the current flow necessary to charge by this mechanism, given that
the observed launch preparation requires typically no more than several seconds.
while still highly resistive, spider silk threads
have a conductivity six orders of magnitude higher than synthetic fibers such a nylon.
The threads also behave as a normal ohmic material except at very low potentials\cite{ES_Thesis2012}.
Based on these measurements, a +250V potential across a meter of
silk (a factor of two above typical ambient vertical potentials) 
would provide enough current to charge at a rate of 10 nC/second, consistent with observed timescales for
ballooning spider launches.

\begin{figure*}[htb!]
\centerline{\includegraphics[width=3.3in]{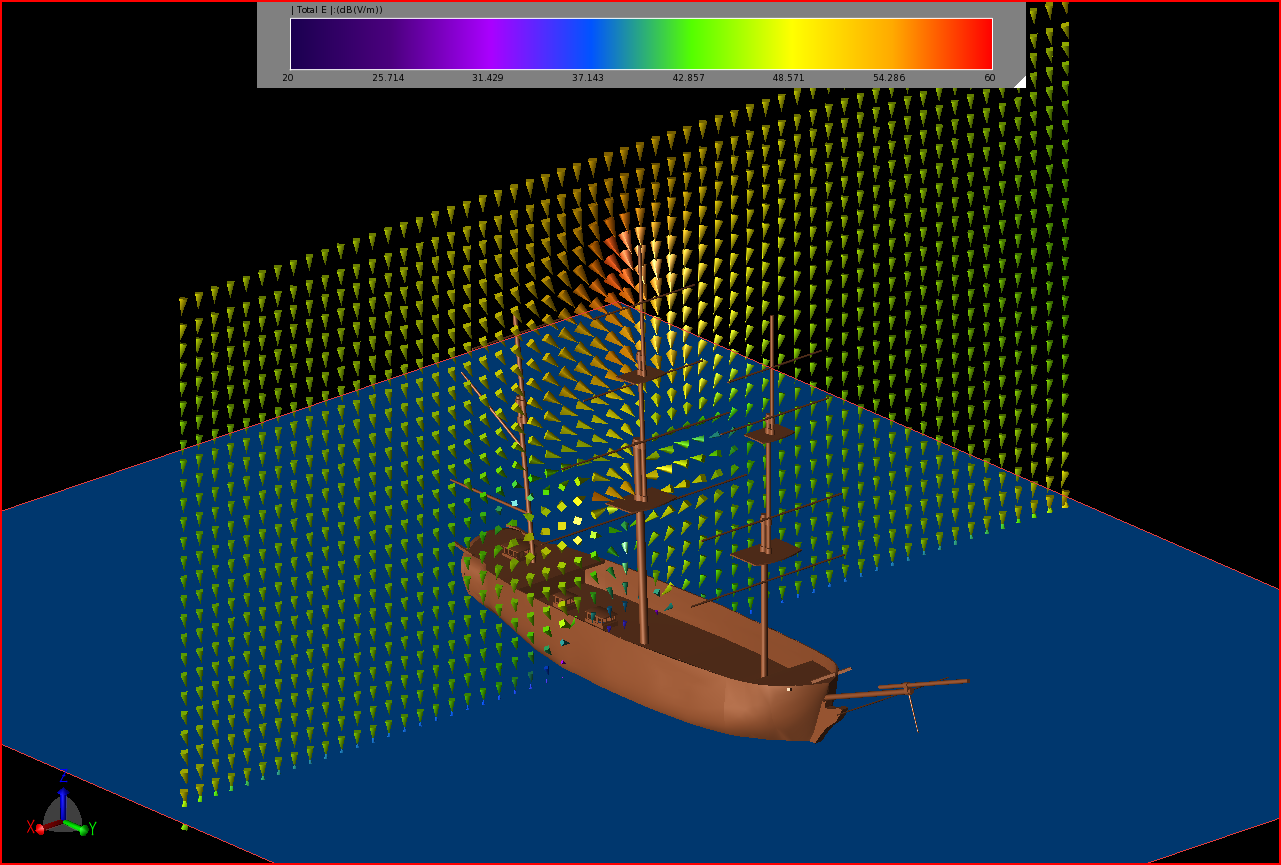}\includegraphics[width=3.52in]{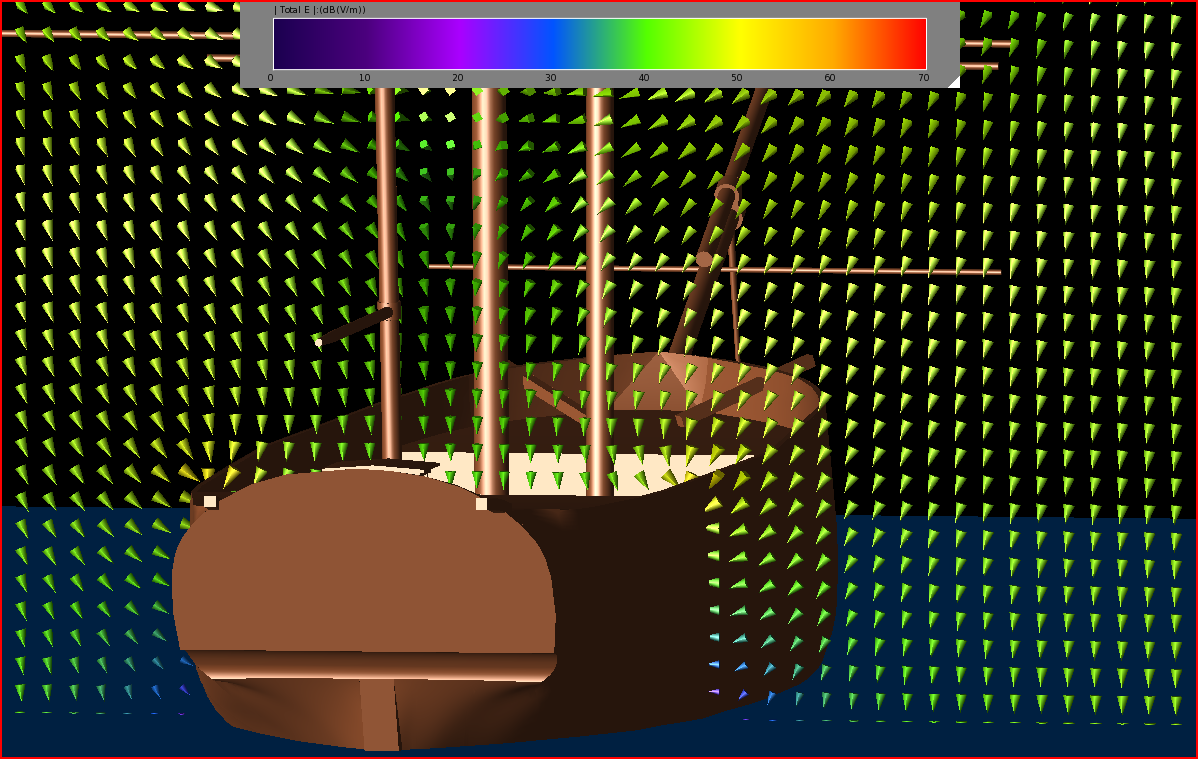}}
\caption{Left: overview of three-dimensional electrostatic field simulation for a computer model
of  HMS Beagle. Vector color is scaled by dB(V/m), and values are set at
an absolute scale similar to the undisturbed Earth's field of 120 V/m. Right:
a zoomed view showing a slice of the electric field vectors in the after-deck
region, indicating the large horizontal component that develops in the vicinity of the ship's
rail. Field strengths are also a factor of 2-3 higher near the rail than in the ambient field.
\label{ShipField}}
\end{figure*}

Regarding the initial acceleration experienced by the spiders, 
and assuming a constant magnitude of acceleration, we may conclude that
for both the Darwin and Schneider {\em et al.}
observations, initial net accelerations in the range of $a_{net} \geq 3$~m~s$^{-2}$
are required. The implied charge is given by
\begin{equation}
Q_{accel} = \frac{m(a_{net}+ g)}{E_0} \simeq 100~{\rm nC}~\left (\frac{m}{1~{\rm mg}}\right ).
\end{equation}
This result is of the same order as the values
required for equilibrium float altitudes of order 100~m or so.

The quantities of charge estimated here, in the tens of nC for even small
spiders, are significant, and the question arises, is there enough capacitance in spider silk
to support charges at this level? For example, such
charge values would imply voltages of 1 kV per nC per pF; 100 nC stored on a conductive wire
could lead to a 100 kV relative potential just due to the stored charge. In this case, the
measured dielectric properties of spider silk are quite important.
Combining these with the cylindrical thread geometry we estimate a capacitance
of order 30 pF per meter. For such values, the implied surface potential, even
for 100 nC per thread, is no more than several kV, far less than (for example) $\sim 10$ kV potentials
that are typical of dry-air charging of macroscopic objects through contact in motion.

One puzzling question remains, however: the statement that the initial
acceleration of the spiders as they left the {\it Beagle} was {\em horizontal} as described by Darwin. 
This seems in sharp tension with the vertical acceleration expected
from the atmospheric potential gradient.
To understand this observation, we used a three-dimensional electrostatic solver to
determine the atmospheric electric field distortion in the
vicinity of a computer-generated model of Darwin's ship.
The ship is assumed to
be at an equipotential with the surface of the ocean. This is appropriate for
the following reasons: (a) the {\it Beagle} carried William Snow Harris' lightning protection
apparatus~\cite{Harris}, including conductors along the mainmast, and these were grounded to seawater;
(b) while oven-dried wood can have very low conductivity, the wood in a ship at sea will
contain a significant amount of moist salt, and under these conditions, conductivity is
many orders of magnitude higher. We estimate that a typical conductivity for the
wood surfaces would be in the range of $10^{-4} - 10^{-3}$~S~m$^{-1}$. Thus while the ship's
surface would not be a good conductor for dynamic currents, under electrostatic conditions we
can expect it to be at equilibrium.

With these assumptions, the results of the field solver~\cite{Solver} are shown in
Fig.~\ref{ShipField}. We find two important details: the field develops a significant horizontal
component near the ships rail over most of its length, and the field strengths close to the
rail are factors of 2-4 higher than the ambient field, reaching several hundred V/m within a
half-meter or so of the ship's surface. 
Given these results, it
appears that the near-horizontal launches observed by Darwin are
consistent with expectations if the charge state of the silk
is relatively high at the time of initial spinning or shortly afterward. Such launches are very difficult
to explain by thermal convection given the calm conditions noted by Darwin.

Regardless of whether the mechanism for the charging of the silk is intrinsic or extrinsic, 
this remarkable behavior -- if it can be confirmed from direct observations of the silk charge state -- 
will place the
Gossamer spider's electrification ability among the most striking evolutionary 
adaptations that Darwin encountered on his voyage. This analysis also highlights 
Darwin's skill and care as an observer: his style of
disinterested observation, accompanied by his detailed notes and narrative, 
stand nearly two centuries later as rich examples for all
exploration and research.

We thank G. Varner and K. Van Houtan for useful comments on the manuscript, and
especially C. Miki for development of the computer model for the Beagle.



\begin{thebibliography}{99}
\bibitem{5strandModel} This 2-D model constrained the total fan included angle to be about 1 radian,
and solved for the strand charge subject to force balance in the Earth's electric field.
A first-order 3-D solution gave similar results.
\bibitem{Solver} The field solver used in these simulations was part of the
Remcom XF7 finite-difference time domain commercial electromagnetic package.
\end{thebibliography}
\end{document}